\documentclass[%
 reprint,
superscriptaddress,
nofootinbib,
 amsmath,amssymb,
prc,
floatfix,
]{revtex4-2}

\usepackage{graphicx}
\usepackage{dcolumn}
\usepackage{bm}
\usepackage[pdftex,colorlinks,linkcolor = blue,urlcolor  = blue,citecolor = blue]{hyperref}

\begin{document}

\preprint{APS/123-QED}

\title{Indirect measurement of the $\pmb{(n,\gamma)^{127}}$Sb cross section}

\author{F.~Pogliano}
\email{francesco.pogliano@fys.uio.no}
\author{A.~C.~Larsen}
\email{a.c.larsen@fys.uio.no}
\author{F.~L.~Bello~Garrote}
\author{M.~M.~Bjørøen}
\author{T.~K.~Eriksen}
\author{D.~Gjestvang}
\author{A.~G\"{o}rgen}
\author{M.~Guttormsen}
\author{K.~C.~W.~Li}
\author{M.~Markova}
\affiliation{Department of Physics, University of Oslo, N-0316 Oslo, Norway}
\author{E.~F.~Matthews}
\affiliation{Department of Nuclear Engineering, University of California, Berkeley, California 94720 U.S.A.}
\author{W.~Paulsen}
\author{L.~G.~Pedersen}
\author{S.~Siem}
\author{T.~Storebakken}
\author{T.~G.~Tornyi}
\author{J.~E.~Vevik}
\affiliation{Department of Physics, University of Oslo, N-0316 Oslo, Norway}

\date{\today}

\begin{abstract}
Nuclei in the $^{135}$I region have been identified as being a possible bottleneck for the \textit{i} process. 
Here we present an indirect measurement for the Maxwellian-averaged cross section of $^{126}\text{Sb}(n,\gamma)$.
The nuclear level density and the $\gamma$-ray strength function of $^{127}$Sb have been extracted from $^{124}$Sn$(\alpha,p\gamma)^{127}$Sb data using the Oslo method. 
The level density in the low-excitation-energy region agrees well with known discrete levels, and the higher-excitation-energy region follows an exponential curve compatible with the constant-temperature model. 
The strength function between  $E_\gamma\approx$ 1.5-8.0 MeV presents several features, such as an upbend and a possibly double-peaked pygmy-like structure.
None of the theoretical models included in the nuclear reaction code TALYS seem to reproduce the experimental data. 
The Maxwellian-averaged cross section for the $^{126}$Sb$(n,\gamma)^{127}$Sb reaction has been experimentally constrained by using our level-density and strength-function data as input to TALYS.
We observe a good agreement with the JINA REACLIB, TENDL, and BRUSLIB libraries, while the ENDF/B-VIII.0 library predicts a significantly higher rate than our results. 
\end{abstract}
\maketitle

\section{Introduction}

The origin of elements heavier than iron in our universe is a hot topic of research among nuclear and astrophysicists, and is regarded as being one of the ``\textit{Eleven Science Questions for the New Century}''~\cite{Questions2003}. 
Since the seminal 
paper of Burbidge \textit{et al.}~\cite{B2HF}, neutron-capture reactions have been identified as the main mechanism for which heavy-element nucleosynthesis take place in stars. 
Now we know that two processes are mainly responsible for the abundances of heavier-than-iron elements in the universe: the $s$ process and the $r$ process, standing for the slow and rapid neutron-capture processes, respectively.
These two processes produce different abundance patterns, and the relative abundances of Ba, La, and Eu on a star's surface may indicate whether the elemental abundance of the star follows an $s$ or $r$ process distribution (see, e.g., ~\cite{ArnouldGoriely2007}). 

One particularly interesting case is the one of carbon-enhanced metal poor stars (CEMPs). These are old stars in the galactic halo and may be enriched in either $r$ process elements~\cite{Hansen_2015}, $s$ process elements, or both~\cite{Sneden}. 
CEMPs enriched in both $s$ and $r$ process elements present a huge challenge. 
Since the two processes are thought to happen in very different astrophysical sites, the mixing of the interstellar medium prior to the formation of the star cannot be the reason behind this peculiar abundance pattern. 
A possible explanation is the presence of an \textit{intermediate} neutron-capture process (the $i$ process) with neutron densities between that of the $s$ and $r$ processes~\cite{Hampel}. 
By assuming that the $i$ process is taking place, both one-zone models and more complex star simulations are able to reasonably reproduce the observed abundances in these stars (see, e.g.,~\cite{Hampel, McKay, Goriely2021}). 
However, all these studies conclude that more accurate estimates of fundamental nuclear properties are needed for a better understanding of the $i$ process.
In particular, neutron-capture rates are of great importance.
Experimental studies of nuclei in the $^{135}$I region are interesting for two reasons. 
First, this region is thought to act as a bottleneck for the $i$ process in CEMP $r$/$s$ stars according to Hampel \textit{et al.}~\cite{Hampel}. 
However, to say how significant this bottleneck might be, one needs information on the neutron-capture rates for the involved nuclei.  
Second, experimental data on fundamental properties of neutron-rich nuclei will help us to develop better and more predictive theoretical models, which both $i$ and $r$ process simulations heavily rely on. 

As neutron-capture rates are extremely hard to measure directly on unstable nuclei, one relies on indirect techniques to constrain these rates. 
At the Oslo Cyclotron Laboratory (OCL), an experimental method has been developed to measure nuclear statistical properties; namely, the $\gamma$-ray strength function (GSF) and the nuclear level density (NLD). 
These two quantities can in turn be used to calculate an experimentally constrained $(n,\gamma)$ cross section (see~\cite{PPNP_Larsen} and references therein).
In this work, we present new data on $^{127}$Sb, produced by the $^{124}$Sn($\alpha,p\gamma$)$^{127}$Sb reaction.
This is the first experiment of a new experimental campaign where neutron-rich nuclei are made by bombarding the most neutron-rich, stable nucleus in an isotopic chain with $\alpha$ particles.
The $^{127}$Sb nucleus is part of the $^{135}$I region, and using our measured  GSF and NLD of $^{127}$Sb we can provide a data-constrained $^{126}$Sb($n,\gamma$)$^{127}$Sb reaction rate for the first time. 

The article is structured as follows: The experimental setup will be described in Sec.~\ref{secExperimental_setup}, and the Oslo method will be presented in Sec.~\ref{secOslo_method}. 
The uncertainty analysis and quantification will be discussed in Sec.~\ref{secUncertainty_analysis} and a discussion on the resulting calculation of the neutron-capture rate in Sec.~\ref{secNCaptureRate}.
Finally, a summary and an outlook are given in Sec.~\ref{secSummary}.

\section{Experimental setup}\label{secExperimental_setup}
The experiment was carried out in November 2020 at the OCL, using an $\alpha$ beam of 24 MeV and $\approx 6$ nA intensity produced by the MC-35 Scanditronix cyclotron. 
The beam impinged on a $^{124}$Sn self-supporting target of 0.47 mg/cm$^2$ thickness and 95.3\% enrichment for a period of six days. 
A short run with a 1-mg/cm$^2$-thick $^{12}$C target was performed for calibration purposes.

\begin{figure}
\includegraphics[width=0.50\textwidth]{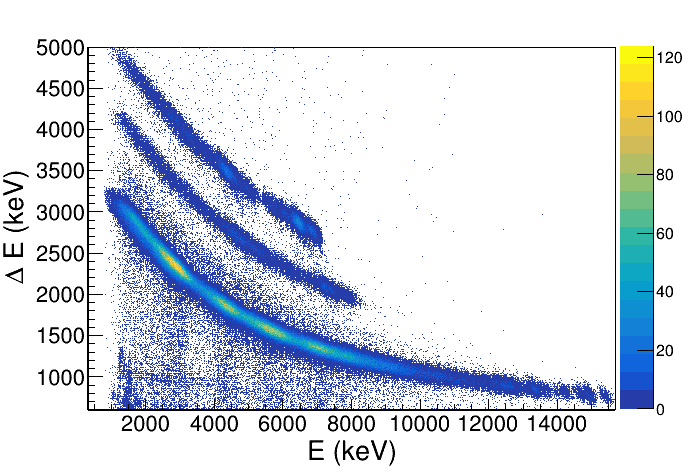}
\caption{\label{fig:bananas} The $\Delta E - E$ plot, or ``banana'' plot, where the energy deposited on the front strip of SiRi ($\Delta E$) is plotted against the one deposited on the back detector ($E$). From bottom to top, we see three bands where the lowest one shows the energies deposited by ejected protons in the $^{124}$Sn($\alpha,p$)$^{127}$Sb reaction, the second deuterons  from the $^{124}$Sn($\alpha,d$)$^{126}$Sb reaction, and the third tritons from the $^{124}$Sn($\alpha,t$)$^{125}$Sb reaction.}
\end{figure}

As we were interested in the particle-$\gamma$ coincidences from the $(\alpha,p \gamma)$ reaction, the Oslo SCintillator ARray (OSCAR) and the Silicon Ring (SiRi) detector arrays were used. 
The targets were placed inside OSCAR~\cite{zeiser2021}, an array of 30 cylindrical (3.5''$\times$8.5'') LaBr3(Ce) $\gamma$-ray detectors mounted on a truncated icosahedron frame. 
The distance between the front of the detectors and the center of the target was 16 cm. 
OSCAR has an energy resolution of 2.7 \% at $E_\gamma = 662$ keV, and fast timing properties with a typical resolution of the prompt timing peak of $\approx 1-5$ ns. 
SiRi~\cite{Guttormsen_NIMA_2011} is a $\Delta E - E$ particle telescope consisting of a ring of eight silicon-telescope modules covering 126$^\circ$-140$^\circ$ in backward angles (corresponding to 6 \% of $4\pi$). 
Each module consists of a thick (1550 $\mu \textrm{m}$) $E$ back detector, with a thin (130 $\mu \textrm{m}$) $\Delta E$ detector in the front. 
The $\Delta E$ detector is divided into eight strips covering about $2^\circ$ each, all together forming a system of 64 detectors. 
To separate the various reaction channels and select only the $(\alpha,p)$ data, we used the $\Delta E - E$ technique, plotting the deposited energy in the back detector versus the deposited energy in the front strip (``banana'' plots; see Fig.~\ref{fig:bananas}). 
The recently installed XiA digital electronics were applied for the data acquisition.

For the energy calibration of OSCAR, we used the 4.439 MeV $\gamma$ transition from the first-excited state of $^{12}$C, together with the 511 keV annihilation peak.
To calibrate SiRi, we used the ground-state peak of $^{127}$Sb in the proton ``banana'' and the  ground-state peak of $^{125}$Sb in the triton ``banana.''

Using the reaction kinematics, we mapped the measured ejectile's energy to excitation energy of the recoiled nucleus, thus providing an excitation energy vs $\gamma$-ray energy 2D spectrum called the ``raw'' coincidence matrix.
Both the excitation energy and $\gamma$-ray energy calibration were then fine-tuned using the known low-lying excited states of $^{127}$Sb, their decay energy, and the nucleus's neutron-separation energy.

\section{The Oslo Method}\label{secOslo_method}

\begin{figure*}
\includegraphics[width=1.0\textwidth]{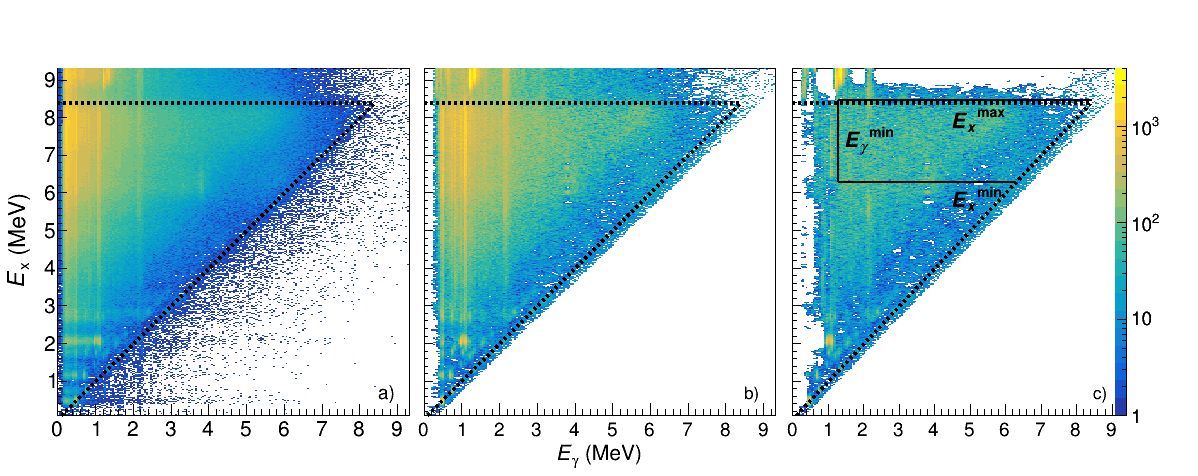}
\caption{\label{fig:3matrices} The (a) raw, (b) unfolded and (c) first-generation matrices used in the Oslo method analysis. On all matrices, the $x$ axis indicates the $\gamma$-ray energy, $E_\gamma$, and the $y$ axis the excitation energy $E_x$. Displayed on all three panels are the $E_x=E_\gamma$ lines and the neutron-separation energy $S_n=8.383$ MeV.}
\end{figure*}

The Oslo method is a set of techniques developed at the OCL to extract the GSF and the NLD from the first-generation $\gamma$-ray matrix~\cite{GUTTORMSEN1996, GUTTORMSEN1987, SCHILLER2000}.
To obtain the first-generation matrix, we start from the calibrated raw matrix shown in Fig.~\ref{fig:3matrices}(a), which must first be unfolded. 
By unfolding, we mean the process of deconvolution; \textit{i.e.}, we estimate the ``true'' signal that was distorted due to the detector response. 
The algorithm is explained in detail in Ref.~\cite{GUTTORMSEN1996}. 
In brief, it is an iterative technique exploiting the fact that folding is a very fast procedure. 
Starting out with a trial function for the ``true'' spectrum, the trial function is folded with the known detector response matrix and compared to the observed spectrum. 
The trial function is then updated accordingly and the process is repeated until good agreement with the observed spectrum is found. 
The unfolding procedure is regularized in two ways: First, the Compton subtraction method is used to preserve the experimental fluctuations bin by bin. 
Second, the ``goodness-of-fit'' is weighted with the experimental fluctuations in addition to the usual $\chi^2$ result. 
Here we use the OSCAR response function~\cite{zeiser2021, fabio_zeiser_2020_4018494}, and the unfolded matrix is presented in Fig.~\ref{fig:3matrices}b. 

An excited nucleus may decay directly to the ground state or go through a $\gamma$-ray cascade, involving one or more lower-lying excited levels, before reaching the ground state. 
To extract the NLD and GSF, we need the \textit{first-generation} (or \textit{primary}) $\gamma$ rays, meaning the first $\gamma$ rays from a cascade. 
These can be extracted by the iterative subtraction method described in Ref.~\cite{GUTTORMSEN1987}.
The main assumption behind this method is that the $\gamma$ spectra are the same whether an excitation-energy bin was populated directly through the reaction, or by $\gamma$ decay from above-lying $E_x$ bins. 
The resulting first-generation matrix is shown in Fig.~\ref{fig:3matrices}c).

The NLD and GSF are average, statistical quantities describing the nucleus in the quasicontinuum, and are the equivalent of levels and reduced transition probabilities in the discrete region.
In the quasicontinuum region, the energy levels are still separable, in principle, as the mean level spacing, $D$, is bigger than the level width, $\Gamma$. 
However, in practice, it is very hard to measure each level and its decay properties, and so it is more useful to describe the nucleus using the NLD and GSF in the quasicontinuum region. 
This excitation-energy region is chosen in the first-generation matrix to extract the NLD and the GSF~\cite{SCHILLER2000}, as shown in Fig.~\ref{fig:3matrices}c). 

The GSF is defined as~\cite{Bartholomew1973}
\begin{equation}
    f^{XL}(E_x,E_\gamma,J,\pi) = \frac{\langle\Gamma_\gamma^{XL}(E_x,E_\gamma,J,\pi)\rangle}{D(E_x,E_\gamma,J,\pi)E_\gamma^{2L+1}},
\end{equation}
where $f^{XL}$ is the GSF for electromagnetic character $X$ and multipolarity $L$ for a transition energy $E_\gamma$, $\left< \Gamma_\gamma^{XL} \right>$ is the average partial $\gamma$ decay width, and $D$ is the mean level spacing. 
In principle, the GSF may depend on excitation energy $E_x$, spin $J$, and parity $\pi$.
The GSF is directly related to the $\gamma$ transmission coefficient by
\begin{equation}\label{Tgeneral}
    \mathcal{T}(E_x,E_\gamma,J,\pi) = \frac{f^{XL}(E_x,E_\gamma,J,\pi)}{2\pi E_\gamma^{2L+1}}.
\end{equation}
The generalized Brink-Axel hypothesis~\cite{Brink, Axel} states that we can average out the dependence on $E_x$, $J$ and $\pi$, allowing us to simplify this expression. 
This hypothesis, central to the Oslo method, is shown to hold for neighboring nuclei of tin~\cite{Masha2021}, and we assume this is the case also for $^{127}$Sb. 
Using this hypothesis, and considering the dipole radiation ($L=1$) to be dominant, we obtain  
\begin{equation}\label{TE1}
    \mathcal{T}(E_\gamma) = \frac{f(E_\gamma)}{2\pi E_\gamma^3}.
\end{equation}
From Fermi's golden rule~\cite{Dirac1927,Fermi1950}, we know that the decay probability is proportional to both the square of 
the matrix element between the initial and final state, and the number of states available in the final excitation-energy bin. 
This is applied in the following ansatz~\cite{SCHILLER2000}:
\begin{equation}\label{PTrhoraw}
    P(E_\gamma,E_x) \propto \mathcal{T}(E_\gamma) \rho(E_x-E_\gamma),
\end{equation}
\textit{i.e.}, the first-generation matrix $P(E_\gamma,E_x)$ is proportional to the product of the two vectors of $\mathcal{T}(E_\gamma)$ and $\rho(E_x-E_\gamma)$, where the latter is the NLD at excitation energy $E_x-E_\gamma$.
This holds as long as we deal with statistical decay: the decay is independent of the way the compound nucleus was originally created.
Therefore, we must make cuts in the first-generation matrix to ensure that this is fulfilled.

For $^{127}$Sb, we choose the following limits:  $E_{x}^{\mathrm{min}}=6.3$~MeV, $E_x^{\mathrm{max}}=8.5$ MeV, and $E_\gamma^{\mathrm{min}}=1.3$ MeV. 
These limits ensure that the excitation energy is high enough for statistical decay to be dominant. The upper $E_x$ limit is just above the neutron-separation energy, so that the spectra are not contaminated with neutron signals. 
The limit on $E_\gamma$ is necessary to prevent the possible inclusion of transitions originating from higher-generation $\gamma$ rays with low transition energies, in particular the strong 1095 keV transition originating from the $(11/2^+)$ level at $E_x = 1095$ keV (see Fig.~\ref{fig:3matrices}a).
Applying these limits on $P(E_\gamma,E_x)$, we estimate the experimental first-generation matrix by~\cite{SCHILLER2000}
\begin{equation}\label{PTrho}
    P(E_\gamma,E_x) = \frac{\mathcal{T}(E_\gamma) \rho(E_x-E_\gamma)}{\sum_{E_\gamma = E_\gamma^{\textrm{min}}}^{E_x}\mathcal{T}(E_\gamma) \rho(E_x-E_\gamma)}.
\end{equation}
The simultaneous extraction of $\mathcal{T}(E_\gamma)$ (and thus $f(E_\gamma)$ by Eq.~(\ref{TE1})) and $\rho(E_x-E_\gamma)$ happens by normalizing the first-generation matrix, $P(E_\gamma,E_x)$, at each excitation energy $E_x$, \textit{i.e.}
\begin{equation}
\sum_{E_\gamma = E_\gamma^{\textrm{min}}}^{E_\gamma} P(E_\gamma,E_x) = 1,    
\end{equation}
and running a $\chi^2$ minimization of Eq.~(\ref{PTrho})~\cite{SCHILLER2000} to extract two  solutions for $\tilde{f}(E_\gamma)$ and $\tilde{\rho}(E_f)$. 
If one solution is found, it can be shown~\cite{SCHILLER2000} that any solution of the form
\begin{subequations}
\begin{align}\label{UnnormRhoFa}
    \tilde{\rho}(E_x - E_\gamma) &= Ae^{\alpha (E_x - E_\gamma)}{\rho}(E_x-E_\gamma)\\
    \tilde{\mathcal{T}}(E_\gamma) &= Be^{\alpha E_\gamma}{\mathcal{T}}(E_\gamma) \label{UnnormRhoFb}
\end{align}
\end{subequations}
also satisfies Eq.~(\ref{PTrho}) for any three parameters \textit{A}, \textit{B} and $\alpha$. 
These are to be determined from experimental data.
Normally, the number of levels at low excitation energies, the $s$-wave level spacing at the neutron-separation energy, $D_0$, and the average total radiative width, $\langle \Gamma_\gamma\rangle$, would be used.
However, such data are typically not available for nuclei far from stability and $^{127}$Sb is no exception. 
We will discuss the normalization of the functions $\rho$ and $\mathcal{T}$ in the following section.

\section{Normalization and uncertainty propagation}\label{secUncertainty_analysis}

\subsection{Normalization of the NLD}

\begin{figure}
\includegraphics[width=0.50\textwidth]{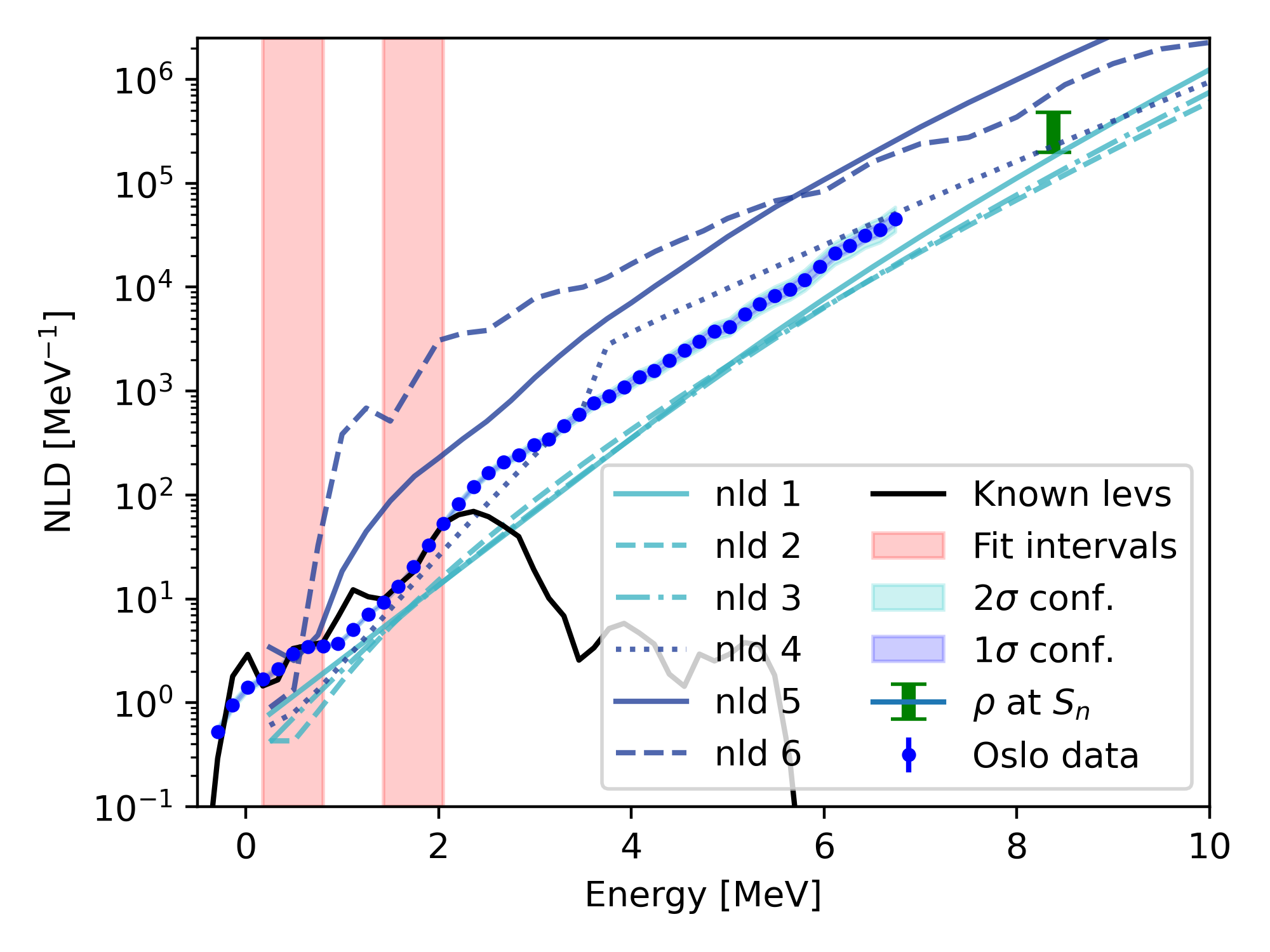}
\caption{\label{fig:nld} Normalization of the NLD (see text) together with the theoretical level density models \texttt{nld} (shorthand for \texttt{ldmodel} 1 to 6) used in TALYS~\cite{TALYS, TALYS2}. The uncertainties in the data points include statistical uncertainties and systematic uncertainties from unfolding and the first-generation method. The total uncertainty band includes also systematic errors from the normalization.}
\end{figure}

The parameters $A$ and $\alpha$ from Eq.~(\ref{UnnormRhoFa}) are needed for the normalization of the NLD.
To constrain these would require two anchor points at low and high excitation energy. 
We normalize our data points in the low-energy region by a fit to known, discrete levels taken from Ref.~\cite{NUDAT}. 
By comparing our data points to the known levels smoothed with our experimental resolution, we observe that the experimental NLD fits very well in the $E_x$ region between $\approx 0.2$ and 0.8 MeV, as well as between $\approx 1.4$ and 2.1 MeV (shaded regions in Fig.~\ref{fig:nld}). 
The apparent ``bump'' in between these intervals could be due to levels in the database that we do not observe in our experiment.
We also normalize our experimental $\rho(E_x)$ to the  level density at the neutron-separation energy. 
This can be calculated from the measured level spacing $D_0$ of \textit{s}-wave neutron resonances at separation energy with~\cite{SCHILLER2000}
\begin{equation}\label{SnfromD0}
    \rho(S_n) = \frac{2\sigma_I^2}{D_0 \left[(I_t + 1)e^{-(I_t+1)^2/2\sigma_I^2} + I_t e^{-I_t^2/2\sigma_I^2}\right]},
\end{equation}
where $I_t$ is the spin of the target nucleus and $\sigma_I$ is the spin-cutoff parameter. 
The use of this formula introduces a model dependence by requiring the estimation of $\sigma_I$ for the spin distribution at the separation energy. 
This is done by assuming a rigid-body moment of inertia:
\begin{equation}\label{eq:RMI}
    \sigma_{I, \textrm{RMI}}^2(S_n) = 0.00146A^{5/3}\frac{1 + \sqrt{1+4aU_n}}{2a},
\end{equation}
where $U_n = S_n - E_{1}$, $E_{1}=-0.45$ MeV is the excitation-energy shift and $a = 12.35$ MeV$^{-1}$ is the level-density parameter, calculated according to the formalism of Ref.~\cite{EgidyBucurescu}. 
The observed experimental values of $\rho(E_x)$ do not reach the separation energy due to the lower limit $E_\gamma^{\mathrm{min}}$, which means the highest $E_x$ is given by $E_x^{\mathrm{max}} -E_\gamma^{\mathrm{min}}$. 
To perform a fit to the $\rho(S_n)$ value, the data must be extrapolated up to $E_x = S_n$. 
This extrapolation introduces another model dependence as one has to assume some model for $\rho(E_x)$ in the gap between our data points and $\rho(S_n)$. 
A commonly used model is the constant-temperature (CT) model, given by the formula~\cite{Ericson,GilbertCameron}
\begin{equation}\label{rho_CT}
    \rho_{\textrm{CT}}(E_x) = \frac{1}{T_{\textrm{CT}}}\exp\left(\frac{E_x-E_0}{T_{\textrm{CT}}}\right),
\end{equation}
where the energy shift, $E_0$, and the nuclear temperature, $T_{\textrm{CT}}$, are parameters to be found from fitting to data. 
Using another model, such as the back-shifted Fermi gas model, gives essentially the same results in the case where the gap between our data and $\rho(S_n)$ is not too large (see Ref.~\cite{Toft_PhysRevC.81.064311}). 

\begin{table*}
\caption{\label{tab:NLD} Values used for the normalization of the NLD. 
The parameters $E_1$ and $a$ are the excitation-energy shift and the level-density parameter, respectively, used in the rigid-body moment of inertia formula in Eq.~(\ref{eq:RMI}). $E_0$ and $T_{\textrm{CT}}$ refer to the parameters used in the constant-temperature model in Eq.~(\ref{rho_CT}), 
while $\rho_{\textrm{f}}(S_n)$ and $\delta{\rho(S_n)}$ represent the limits for which the level density at neutron-separation energy is flat, and the width of the tapering outside these limits, respectively. Finally, $\sigma_I$ is the spin-cutoff parameter calculated by Eq.~(\ref{eq:RMI}) and $D_0$ is the range of level spacings of \textit{s}-wave neutron resonances related to $\rho_f(S_n)$ by Eq.~(\ref{SnfromD0}).}
\begin{ruledtabular}
\begin{tabular}{ccccccccc}
$E_1$ & $a$ & $E_0$ & $T_{\textrm{CT}}$ & $\rho_{\textrm{EB}}(S_n)$ & $\rho_{\textrm{f}}(S_n)$  & $\delta{\rho(S_n)}$ & $\sigma_I$ & $D_0$\\ 
 (MeV) & (MeV$^{-1}$) & (MeV) & (MeV) & (10$^3$ MeV$^{-1}$) & (10$^3$ MeV$^{-1}$) & (10$^3$ MeV$^{-1}$) & & (eV)\\ \hline
-0.45 & 12.35 & (-1.2,-1.9)\footnote{Varying according to the choice of $\rho_f(S_n)$} & 0.8 & 376 & 199$-$481& 90& 6.45 & 58.8$-$24.3\\
\end{tabular}
\end{ruledtabular}
\end{table*}

Experimental $D_0$ values are typically available for stable nuclei, from which $\rho(S_n)$ can be derived using Eq.~(\ref{SnfromD0}). 
For unstable nuclei, the value of $\rho(S_n)$ must be obtained by other means. 
In this work, we compare theoretical values to the semiexperimental values for nuclei in the same mass region as $^{127}$Sb. 
Thus, we apply a similar strategy to the one in Kullmann \textit{et al.}~\cite{Ina191Os}, where $D_0$ values for the neighboring isotopes of Sn, Sb and Te (corresponding to $Z=50,51$ and 52, respectively) from both the \textit{Atlas of Neutron Resonances}~\cite{MughabghabS.F.2018Aonr} and the \textit{Reference Input Parameter Library}~\cite{RIPL} were used to calculate $\rho(S_n)$. 
These values are then compared to the theoretical $\rho(S_n)$ estimates using the global parametrization of Ref.~\cite{EgidyBucurescu} to evaluate how well they agree. 
From this, $\rho_{\textrm{EB}}(S_n)=376 \times 10^3\textrm{ MeV}^{-1}$ is obtained for $^{127}$Sb.

A conservative estimate of the uncertainty of $\rho(S_n)$ for $^{127}$Sb from this evaluation would be a flat probability distribution between $0.53\rho_{\textrm{EB}}$ and $1.28\rho_{\textrm{EB}}$, 
where the edges of the distribution are smoothed with a Gaussian with a standard deviation of $\delta\rho = 90 \times 10^3$ MeV$^{-1}$. 
The probability distribution was chosen to be flat between the two values $0.53\rho_{\textrm{EB}}$ and $1.28\rho_{\textrm{EB}}$, 
as there is no clear reason to prefer one value over another within this range.
The value of $\delta\rho$ is not straightforward to obtain. 
However, we believe that we have chosen a reasonable estimate, as it corresponds to what is obtained by translating the uncertainty in the value of $D_0$ for neighboring nuclei to $\rho(S_n)$ using Eq.~(\ref{SnfromD0}). See Table~\ref{tab:NLD} for an overview of all the parameters used in the NLD normalization.

Many different NLD normalizations were generated with the \texttt{counting.c} code from the Oslo method software~\cite{OsloMethodSoftware} by varying the input parameters to the code.
The code normalizes the experimental NLD by running a $\chi^2$ minimization of the (unnormalized) experimental data fitting it to the known levels at low $E_x$, and to the CT model that goes through $\rho(S_n)$ at high $E_x$. 
The input parameters that are changed are the lower- and upper-energy bin constraining the fitting interval (L1 and L2) for the $\chi^2$ minimization in the low-$E_x$ region.
Further, the $\rho_{\textrm{CT}}$ formula in Eq.~(\ref{rho_CT}) is used to interpolate the level density between our data points and  $\rho(S_n)$. 
The parameters $E_0$ and $T_{\textrm{CT}}$ from Eq.~(\ref{rho_CT}) are determined in \texttt{counting.c} by providing a fitting interval to the data at high $E_x$ and to the value of $\rho(S_n)$. 

Above a given $E_x$, the level density becomes more smooth and the information about the known levels starts to become incomplete (in this case, where $E_x\gtrsim 3$ MeV). 
As the CT model is essentially an exponential function, and our data points display a very smooth trend where $E_x \approx 3-6.5$ MeV, the $T_{\textrm{CT}}$ parameter was found to vary very little when choosing different data points for the fit. 
Therefore, these fitting points were kept fixed, using a temperature parameter of $T_{\textrm{CT}} = 0.8$ MeV. 
The $E_0$ parameter in the CT formula is the shift parameter, and for a fixed $T_{\textrm{CT}}$, this will change according to the choice of $\rho(S_n)$. 
The $E_0$  parameter was found to have values between $E_0=-1.2$ MeV (for 0.53$\rho_{\textrm{EB}}$) and $E_0=-1.9$ MeV (for 1.28$\rho_{\textrm{EB}}$.)

\begin{figure}
\includegraphics[width=0.50\textwidth]{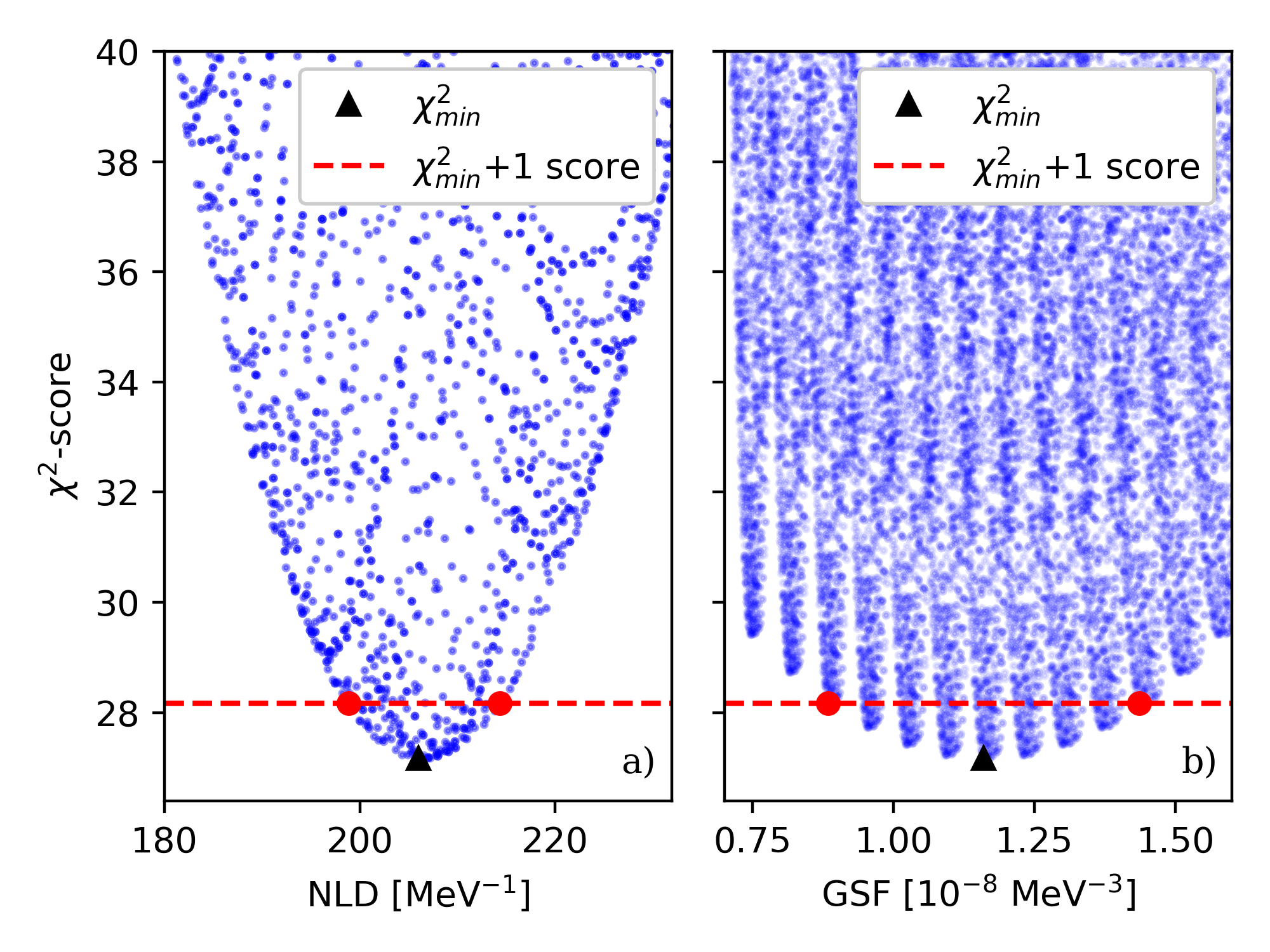}
\caption{\label{fig:chi2s} The $\chi^2$ scores of each calculated NLD (a) and GSF (b) for $E_x=2.68$ MeV and $E_\gamma=2.68$ MeV. Each $E_x$ and $E_\gamma$ bin has a similar, parabola-shaped distribution of $\chi^2$-scores. 
From these we estimate the uncertainty of every bin by checking graphically where the parabola crosses the $\chi^2+1$ line (red points).
The mean value is where $\chi^2=\chi^2_{min}$ (black triangles).
}
\end{figure}

The \texttt{counting.c} code was run with every L1 and L2 combination so that $\mathrm{L}1<\mathrm{L}2\leq22$ (where bin 22 corresponds to $E_x=2.68$ MeV), and a range of 50 values for $\rho(S_n)$ between $0.4\rho_{\textrm{EB}}$ and $1.4\rho_{\textrm{EB}}$, incorporating the smoothing of the edges as mentioned above. 
This range corresponds to 11500 different parameter combinations, and thus differently normalized NLDs. 
For each of these combinations, we calculate a $\chi^2_{\textrm{NLD}}$ score through
\begin{equation}
    \chi^2_{\textrm{NLD}} = \sum_i \frac{(\rho_{n}(E_i) - \rho_{k}(E_i))^2}{\Delta\rho_{n}(E_i)^2} + \chi^2_{S_n},
\end{equation}
where the sum runs over the energy bins $i = \{6,7,8,9,10,14,15,16,17,18\}$, where the results seem to agree the most with the known levels and are shown in the shaded regions of Fig.~\ref{fig:nld}. $\rho_{n}(E_i)$ is the value at the \textit{i}th energy bin of the normalized NLDs, $\Delta\rho_{n}(E_i)$ its associated statistical uncertainty from the experiment, and $\rho_{k}(E_i)$ the level density calculated from the smoothed known levels. 
Finally, $\chi^2_{S_n}$ keeps track of the uncertainty for the normalization tied to the choice of $\rho(S_n)$ and is calculated as
\begin{equation}
    \chi^2_{S_n}=
    \begin{cases}
    \frac{\left(\rho_{n}(S_n) - 0.53\rho_{\textrm{EB}}\right)^2}{\rho_\sigma^2}, & \text{if}\ \rho_{n}(S_n)/\rho_{\textrm{EB}} < 0.53 \\
    0, & \text{if}\ 0.53 \leq \rho_{n}(S_n)/\rho_{\textrm{EB}} \leq 1.28 \\
    \frac{\left(\rho_{n}(S_n) - 1.28\rho_{\textrm{EB}}\right)^2}{\rho_\sigma^2}, & \text{if}\ \rho_{n}(S_n)/\rho_{\textrm{EB}} > 1.28,
    \end{cases}
\end{equation}
mimicking what a $\chi^2$ score would look like for a flat distribution inside a range, and otherwise behaving as a normal distribution.
With 11500 different NLDs [and thus 11500 $\rho(E_x)$ values for each energy bin], each with its own $\chi^2_{\textrm{NLD}}$, we are able to find the mean value of each bin by choosing the value with the smallest $\chi^2_{\textrm{NLD}}$. 
Then, for each $E_i$ bin, the uncertainty was graphically determined by checking where the parabola-like $\chi^2_{\textrm{NLD}}(\rho(E_i))$ plot crossed the $\chi^2_{min}+1$ line (see Fig.~\ref{fig:chi2s}a).

We note that the NLD normalization is rather strongly constrained by the fit to the known levels in the two shaded regions shown in Fig.~\ref{fig:nld}. 
Therefore, despite the large uncertainty in the normalization point $\rho_{\mathrm{EB}}(S_n)$, the slope of the NLD data points (and thus the slope of the GSF) is quite well determined.

In Fig.~\ref{fig:nld}, together with the experimental results, we show the six NLD models included in TALYS 1.95~\cite{TALYS,TALYS2}. 
Here \texttt{ldmodel} 1 combines a constant-temperature model with the Fermi
gas model \cite{GilbertCameron}, \texttt{ldmodel} 2 is the back-shifted Fermi gas model \cite{GilbertCameron}, \texttt{ldmodel} 3 the generalized superfluid model \cite{ignatyuk79, ignatyuk93}, and \texttt{ldmodel} 4-6 are Hartree-Fock-based calculations. 
From Fig.~\ref{fig:nld} we can observe that most of the models fail at reproducing the experimental results, and do not even meet the conservative error estimate for $\rho(S_n)$, with the exception of \texttt{ldmodel} 4 that comes the closest to the Oslo data.

\subsection{GSF}
\begin{figure}
\includegraphics[width=0.50\textwidth]{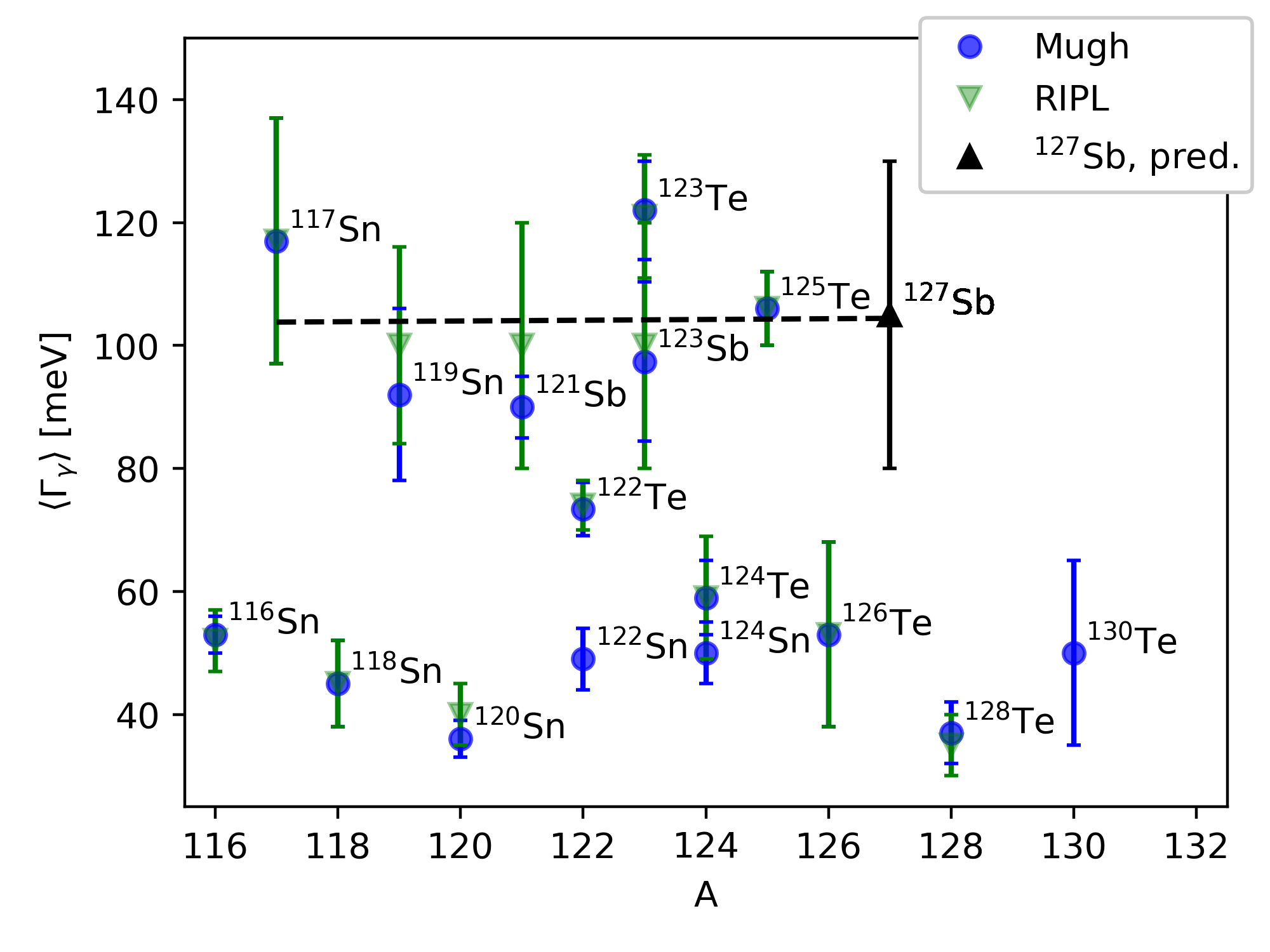}
\caption{\label{fig:Gg_syst} Values of  $\langle \Gamma_\gamma \rangle$  from Mughabghab~\cite{MughabghabS.F.2018Aonr} and RIPL~\cite{RIPL} for the neighboring nuclei of $^{127}$Sb. The black dashed line indicates the linear regression for the $\langle \Gamma_\gamma \rangle$ values of the odd-even nuclei.}
\end{figure}

The last free parameter in Eq.~(\ref{UnnormRhoFb}) is $B$, responsible for the absolute normalization of the GSF. 
Since the average total radiative width, $\langle \Gamma_\gamma \rangle$, for $^{127}$Sb is not known, it is again necessary to use systematics from neighboring nuclei to assess its value. 
In Fig.~\ref{fig:Gg_syst} we show that the $\langle \Gamma_\gamma \rangle$ values for different nuclei in this mass region, gathered from Ref.~\cite{MughabghabS.F.2018Aonr}. 
Two patterns are observed, one for the even-even and one for the odd-even nuclei, respectively. 
As $^{127}$Sb is odd-even, we use the data from the other odd-$A$ nuclei to estimate $\langle \Gamma_\gamma \rangle$. 
Either an average or an extrapolation from linear regression could be used to predict the $\langle \Gamma_\gamma \rangle$ value for $^{127}$Sb, and fortunately, both yield about the same values, rounded to $\langle \Gamma_\gamma \rangle_\mu=105$ meV. 
The uncertainty is taken to be normally distributed and is kept conservatively to be $\langle \Gamma_\gamma \rangle_\sigma = 25$ meV (see Fig.~\ref{fig:Gg_syst}).

\begin{figure}
\includegraphics[width=0.50\textwidth]{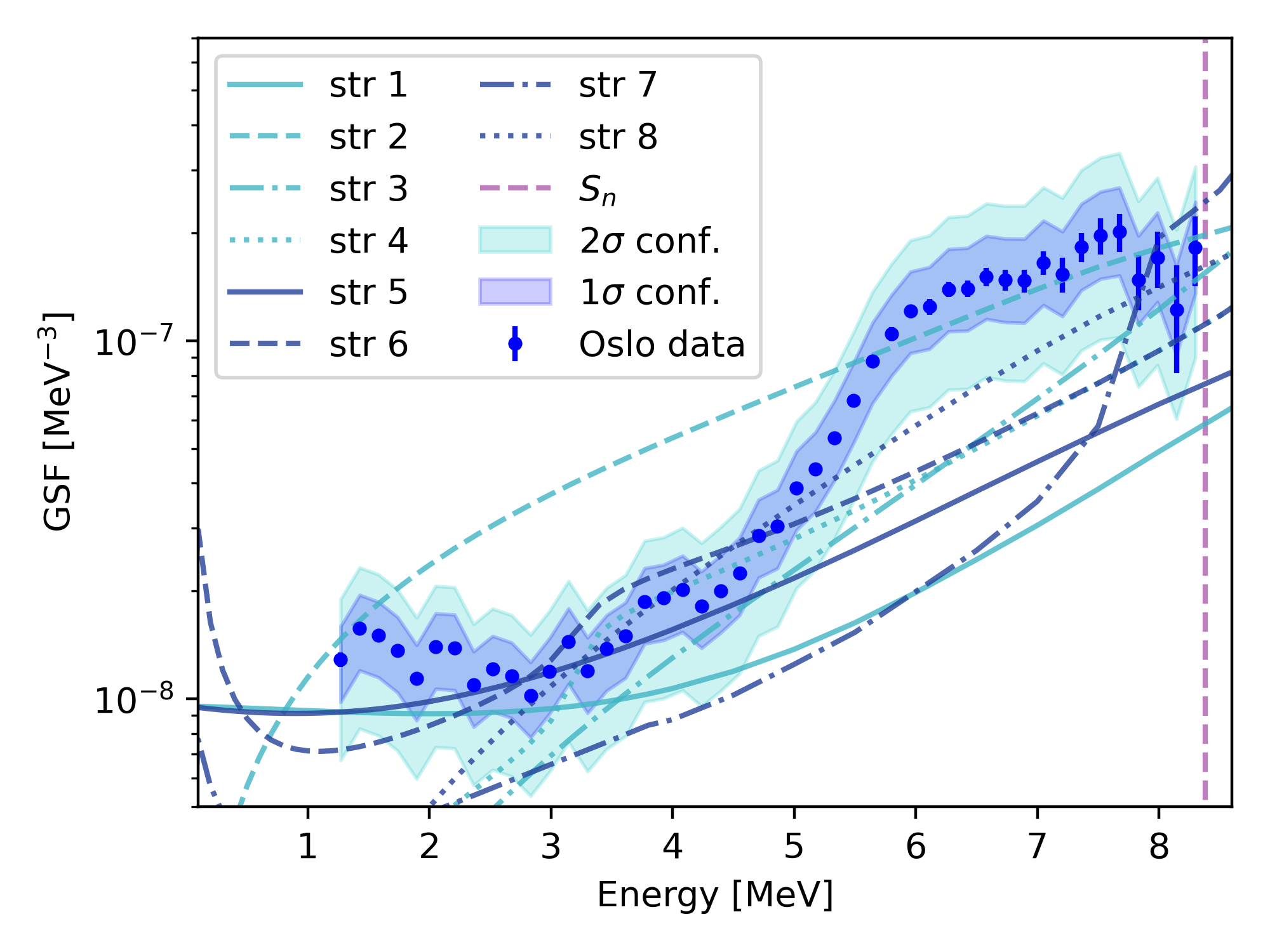}
\caption{\label{fig:gsf} The normalized GSF, together with the theoretical models \texttt{str} (shorthand for \texttt{strength} 1 to 8) used in TALYS~\cite{TALYS, TALYS2}. The uncertainties in the data points include statistical uncertainties and systematic uncertainties from unfolding and the first-generation method. The total uncertainty band includes also systematic errors from the normalization.}
\end{figure}

To get the absolute normalization of the GSF, we use the script \texttt{normalization.c}~\cite{OsloMethodSoftware}, that takes as input the $\langle \Gamma_\gamma \rangle$ value, the estimated $D_0$ value which (with the given spin distribution)  reproduces $\rho(S_n)$ used for the NLD data points, the normalized NLD, and the $\gamma$-ray transmission coefficient normalized in slope with the  parameter $\alpha$. 
By choosing 13 different $\langle \Gamma_\gamma \rangle$ values between $\langle \Gamma_\gamma \rangle = 65.5-142.5$ meV, we  run \texttt{normalization.c} for each NLD obtained from \texttt{counting.c}. 
This gives us 149500 different GSFs, 13 for each NLD.
Each GSF inherits the $\chi^2_{\textrm{NLD}}$ score from the associated NLD, and to assess the ``goodness-of-fit'' we add a term accounting for the deviation of the chosen $\langle \Gamma_\gamma \rangle_n$ from the mean value $\langle \Gamma_\gamma \rangle_\mu = 105$ meV:
\begin{equation}
    \chi^2_{\textrm{GSF}} = \chi^2_{\textrm{NLD}} + \frac{\left(\langle \Gamma_\gamma \rangle_n - \langle \Gamma_\gamma \rangle_\mu \right)^2}{\langle \Gamma_\gamma \rangle_\sigma^2}.
\end{equation}
Similarly to the NLD calculations, the GSF evaluated at each $E_\gamma$ energy bin will have a mean value corresponding to where $\chi^2_{\textrm{GSF}}=\chi^2_{\textrm{min}}$, and an uncertainty where the $\chi^2_{\textrm{min}}(f(E_\gamma))+1$ line crosses the parabola. 
This is shown for one specific bin ($i = 22$, where $E_i=2.68$ MeV) in Fig.~\ref{fig:chi2s}b. 
The resulting GSF with the corresponding errors is displayed in Fig.~\ref{fig:gsf}.

\subsection{Comparison with GSF models}
We observe that the GSF shows similar features to the ones found in the neighboring nuclei - most prominently, a resonance-like peak at about $E_\gamma \approx 7$ MeV and a low-energy enhancement below $E_\gamma \approx 3$ MeV.
These structures can be interpreted as the pygmy dipole resonance (PDR) and the upbend on top of the tail of the giant dipole resonance (GDR). 
Along with the experimental results, Fig.~\ref{fig:gsf} shows the eight theoretical GSF models available from the reaction code TALYS~1.95~\cite{TALYS,TALYS2}. 
None of these seem to fit well, as none predict such a strong pygmy-like structure as observed in the data. 
Although the upbend is included in four of them (\texttt{strength} 1, 5, 6 and 7), none seem to give a fully correct behavior. 
The GFSs modeled by \texttt{strength} 1 and 2 are the generalized Lorentzian model \cite{Kopecky_PhysRevC.41.1941} and the standard Lorentzian (Brink-Axel) model \cite{BRINK57,Axel}, respectively. 
These are phenomenological models, and are not expected to give good predictions for nuclei far from stability. 
All other models are microscopic, mostly based on the quasiparticle random-phase approximation. These models attempt to describe nuclei from the underlying physics rather than by phenomenology. 
However, none of them manage to predict the PDR for $^{127}$Sb in a satisfying way; they all systematically underestimate the strength in the $E_\gamma = 5-7$ MeV region. 
This underestimate of the PDR may consequently lead to systematic underestimates of $(n,\gamma)$ rates used in astrophysical applications.

The main feature of the GSF for transition energies below $S_n$ is the tail of the GDR, and also the PDR. 
The GDR tail can be modeled by a generalized Lorentzian (GLO)~\cite{Kopecky_PhysRevC.41.1941},
\begin{multline}\label{eq:GLO}
    f^{\textrm{GLO}}(E_\gamma) = \frac{\sigma_0\Gamma_0}{3\pi^2 \hbar^2 c^2}\times\\\left(\frac{E_\gamma\Gamma_K}{\left(E^2_\gamma - E^2_0\right)^2 + E_\gamma^2 \Gamma^2_K}+0.7\frac{\Gamma_{K,0}}{E^3_\gamma}\right),
\end{multline}
where
\begin{equation}
    \Gamma_K(E_\gamma,T_f) = \frac{\Gamma_0}{E^2_0}\left(E^2_\gamma + 4\pi^2T_f^2\right)
\end{equation}
and $\Gamma_{K,0} = \Gamma_K(0,T_f)$.
$\sigma_0$, $E_0$, $\Gamma_0$ and $T_f$ are considered free parameters, representing the peak cross section, the energy centroid, the width and the temperature of the final levels, respectively. 
As there are no experimental photonuclear data of $^{127}$Sb, we  infer the GLO parameters by again comparing to data from neighboring nuclei. 
We choose the GLO parameters by averaging over the values of fitting the GDRs for $^{126}$Sn and $^{128}$Te, these being the nuclei directly below and above $^{127}$Sb in the nuclear chart, respectively. 
As a proxy for the GDR of $^{128}$Te, we sum over the $^{128}$Te($\gamma,n$) and $^{128}$Te($\gamma,2n$) cross sections from Lepretre \textit{et al.}~\cite{128Te}, while the GLO parameter values of $^{126}$Sn are found by extrapolation from the lighter isotopes of tin~\cite{Toft_tins, Darmstadt_tins, Masha2021}. 
These two approaches give GLOs quite similar in magnitude and shape, and we estimate the GLO parameters of $\sigma_0$, $E_0$, and $\Gamma_0$, for $^{127}$Sb to be the mean of the corresponding values found for $^{126}$Sn and $^{128}$Te. 
However, in order to find an appropriate value for $T_f$, we need information on the low-energy tail (well below $S_n$), which is not available from photonuclear data as the data from Lepretre \textit{et al.}~\cite{128Te} only probes the GSF from $S_n$ and higher $E_\gamma$. 
We choose to use the same $T_f$ as applied for the tin isotopes, with a large  uncertainty. 
All parameters can be found in Table~\ref{tab:127Sb_GDR_fit}.

\begin{table}
\caption{\label{tab:127Sb_GDR_fit} Fitting parameters for the GLO of $^{126}$Sn, $^{127}$Sb and $^{128}$Te (see text).}
\begin{ruledtabular}
\begin{tabular}{ccccc}
Nucleus     & $T_f$ & $E_0$ & $\Gamma_0$ & $\sigma_0$ \\
&(MeV) & (MeV) & (MeV) & (mb) \\ \hline \rule{0pt}{2.5ex}
 $^{126}$Sn & $0.30(10)$ & $15.3(3)$ & $4.6(6)$ & $265(22)$ \\
 $^{127}$Sb & $0.30(30)$\footnote{Estimated from tin isotopes.}
 & $15.4(4)$ & $5.4(10)$ & $283(28)$ \\
 $^{128}$Te & $0.30(30)^a$ & $15.4(1)$ & $6.1(4)$ & $301(5)$ \\ 
\end{tabular}
\end{ruledtabular}
\end{table}

\begin{figure}
\includegraphics[width=0.50\textwidth]{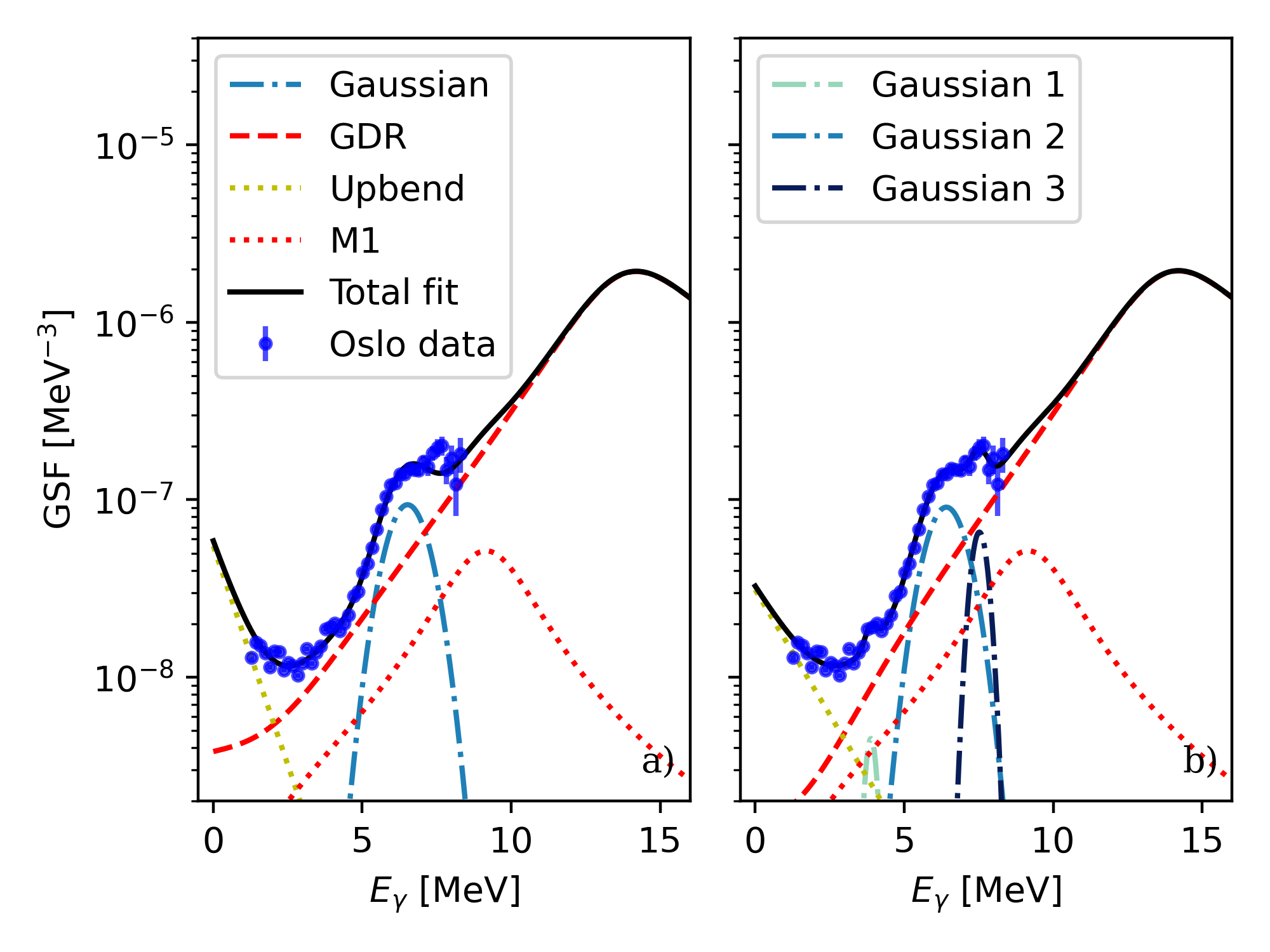}
\caption{\label{fig:gsf_fit} Two possible decompositions of the GSF with one (a) and two (b) Gaussians for the PDR region (see discussion in text). The uncertainties in the data points correspond to the statistical uncertainty and systematic uncertainties from unfolding and the first-generation method.}
\end{figure}

Figure~\ref{fig:gsf_fit} shows two different attempts to decompose the GSF into its constituent structures. 
The GDR, the upbend and a pygmy-like structure at $\approx 7$ MeV, and the spin-flip M1 resonances are included in both. 
The GDR was modeled with a GLO using the parameters in Table \ref{tab:127Sb_GDR_fit}, the upbend by an exponential function of the form
\begin{equation}\label{eq:upbend}
    f^{\textrm{up}}(E_\gamma) = C_{\textrm{up}}e^{-a_{\textrm{up}}E_\gamma}
\end{equation}
and the spin-flip M1 resonances by a standard Lorentzian
\begin{equation}\label{eq:SLO}
    f^{\textrm{SLO}}(E_\gamma) = \frac{1}{3\pi^2 \hbar^2 c^2}\frac{\sigma_s\Gamma_s^2 E_\gamma}{\left(E^2_\gamma - E^2_s\right)^2 + E_\gamma^2 \Gamma^2_s},
\end{equation}
where $\sigma_s$, $\Gamma_s$, and $E_s$ are free parameters and correspond to the same quantities as for the GLO in Eq.~(\ref{eq:GLO}). These parameters were determined by extrapolation of the fittings of the M1 strengths measured in lighter tin isotopes \cite{Darmstadt_tins} similarly as what done with the GDR.
In Fig.~\ref{fig:gsf_fit}a, a single Gaussian is employed to describe the pygmy structure, while in Fig.~\ref{fig:gsf_fit}b two Gaussians are employed. The choice of employing Gaussians to model resonances is unorthodox, but gives a better fit than using a more conventional Lorentzian when applied to the PDR. This is as also observed for other nuclei (\textit{e.g.}, tin isotopes in Ref.~\cite{Masha_master}) and the reason for this is unknown. The Gaussian function is given by
\begin{equation}
    f^{\mathrm{Gauss}}(E_\gamma) = \frac{1}{\sqrt{2\pi}\sigma_r}\cdot C \exp{\left[ -\frac{(E_\gamma - E_{r})^2}{2\sigma_r^2}\right]},
\end{equation}
where $\sigma_r$ is the standard deviation, $C$ is a normalization constant, and $E_r$ is the centroid (expected value). 
A satisfying fit of the PDR is obtained with only one Gaussian function, although the data points at the highest $E_\gamma$  are not fully reproduced.
This fit gives an integrated, energy-weighted cross section of  $\approx 0.7$\% of the Thomas-Reiche-Kuhn energy-weighted sum rule (EWSR)~\cite{Thomas1925, Reiche1925, Kuhn1925}, see Table~\ref{tab:GSF_Gauss_components_a}.
Three Gaussians are needed to reproduce all the visible structures as shown in Fig.~\ref{fig:gsf_fit}b, yielding $\approx 0.9$\% of the sum rule, see Table~\ref{tab:GSF_Gauss_components_b}. 
While the use of two Gaussians has been done in Ref.~\cite{Masha_master} to describe the PDR in tin isotopes, an additional, smaller structure is observed for $^{127}$Sb at $E_\gamma \approx 4$ MeV.
A similar feature is present in the $^{117}$Sn GSF at $E_\gamma \approx 2.5$~MeV~\cite{Agvaanluvsan2009,Toft_tins} and they might have the same origin.
Although the energy region could coincide with that of the scissors mode, both tin and antimony with their respective proton numbers of 50 and 51 are known to be almost spherical nuclei, while the scissors mode is observed only in deformed nuclei.

Although it was found that the integrated, energy-weighted cross section of the pygmy-like structure is $\approx 0.8$\% of the EWSR, it should be emphasized that this is a conservative estimate. 
In this work a fitted GLO ``background'' with a maximal $E1$ strength is employed. 
Considering that theoretical models (\textit{e.g.} those in TALYS) give a rather low GDR tail (see Fig.~\ref{fig:gsf}), the fraction could be considerably larger.

\begin{table}
\caption{\label{tab:GSF_Gauss_components_a} Parameters used for the  Gaussian in Fig.~\ref{fig:gsf_fit}a. 
}
\begin{ruledtabular}
\begin{tabular}{lcccc}
Function & $E_r$ & $\sigma_r$ & $C$ & EWSR \\ 
 & (MeV) & (MeV) & ($10^{-9}$ MeV$^{-2}$) &  (\%) \\\hline \rule{0pt}{2.5ex} 
 Gauss   & $6.52(5)$ & $0.70(3)$ & $164(9)$ & $0.72(5)$ \\
\end{tabular}
\end{ruledtabular}
\end{table}

\begin{table}
\caption{\label{tab:GSF_Gauss_components_b} Parameters used for the three Gaussians in Fig.~\ref{fig:gsf_fit}b. The EWSR  is calculated only for the second and third Gaussian, fitting the pygmy-like peak. }
\begin{ruledtabular}
\begin{tabular}{lcccc}
Function & $E_r$    & $\sigma_r$ & $C$ & EWSR  \\ 
         & (MeV)    &  (MeV)   &  ($10^{-9}$ MeV$^{-2}$) & (\%) \\ 
\hline
Gauss1   & $3.88(2)$ & $0.19(3)$ & $2.2(4)$             & - \\
Gauss2   & $6.41(7)$ & $0.69(4)$ & $157(15)$             & - \\
Gauss3   & $7.52(8)$ & $0.28(9)$ & $46(15)$             & - \\ \hline
 Sum      & -              & - & - & $0.9(2)$\footnote{Calculated only for the last two Gaussians.} \\
\end{tabular}
\end{ruledtabular}
\end{table}

\section{Neutron-capture cross section}\label{secNCaptureRate}
The radiative neutron-capture rate [or $(n,\gamma)$-rate] and the Maxwellian-averaged cross section (or MACS) are of particular interest for astrophysical applications such as nucleosynthesis network calculations. 
These quantities are closely related by \cite{Iliadis}
\begin{equation}
    N_A \langle \sigma v \rangle = \frac{N_A \langle \sigma \rangle}{v_T},
\end{equation}
where $N_A \langle \sigma v \rangle $ is the $(n,\gamma)$-rate, $N_A \langle \sigma \rangle$ the MACS, $N_A$ is Avogadro's number, and $v_T=\sqrt{2k_B T /\tilde{m}}$ is the thermal speed.
Further, $k_B$, $T$, and $\tilde{m}$ are the Boltzmann constant, the temperature, and the reduced mass of the neutron plus the target nucleus, respectively. 
The $(n,\gamma)$ rate can then be calculated by (see,  e.g.,~\cite{ArnouldGoriely2007})
\begin{multline}\label{ncrate}
    N_A\langle\sigma v\rangle(T)=\left(\frac{8}{\pi \tilde{m}}\right)^{1/2}\frac{N_A}{\left(k_B T\right)^{3/2}G_t(T)} \times \\\int_0^\infty \sum_\mu \frac{2J_t^\mu + 1}{2J_t^0 + 1}\sigma^\mu_{n\gamma}(E)E\exp\left[-\frac{E+E_x^\mu}{k_B T}\right]\textrm{d}E,
\end{multline}
where $J_t^0$ and $J_t^\mu$ are the spin for the ground state and the $\mu$th excited state respectively, $E_x^\mu$ the energy of the $\mu$th excited state, $E$ the relative energy between the neutron and the target nucleus, $\sigma^\mu_{n\gamma}$ the $(n,\gamma)$ cross section for the target nucleus excited to the $\mu$'th state, and $G_t(T)$ is the partition function given by
\begin{equation}
    G_t(T) = \sum_\mu \frac{2J_t^\mu + 1}{2J_t^0 + 1}\exp\left[\frac{-E_x^\mu}{k_B T}\right].
\end{equation}

The radiative neutron-capture cross section ($\sigma^\mu_{n\gamma}$) in Eq.~(\ref{ncrate}) can be calculated from either theoretical or experimental values of the NLD and GSF for the compound nucleus in the Hauser-Feschbach framework~\cite{RAUSCHERTHIELEMANN}. 
Recommended theoretical values for either the $(n, \gamma)$ rate, the MACS, or both can be found in libraries such as the JINA REACLIB rates~\cite{JINAREACLIB}, TENDL-19~\cite{TENDL-19}, BRUSLIB~\cite{BRUSLIB} and ENDF/B-VIII.0~\cite{ENDF8}. 

\begin{figure}
\includegraphics[width=0.50\textwidth]{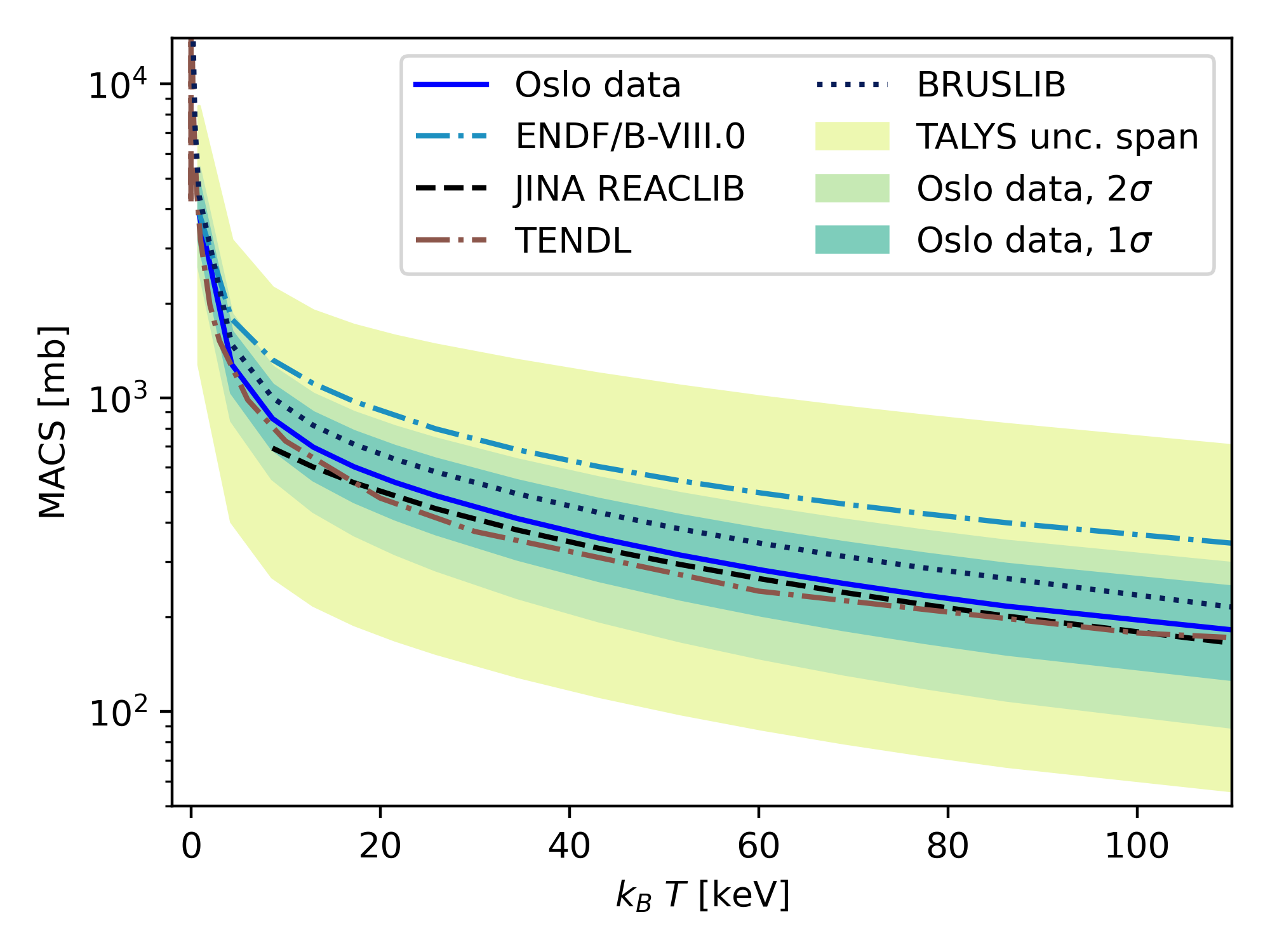}
\caption{\label{fig:MACSs} The calculated experimentally constrained MACS for the $^{126}$Sb$(n,\gamma)$ reaction, together with theoretical values from JINA REACLIB~\cite{JINAREACLIB}, TENDL~\cite{TENDL-19}, BRUSLIB~\cite{BRUSLIB} and ENDF/B-VIII.0~\cite{ENDF8}.}
\end{figure}

From the experimentally constrained NLD and GSF of $^{127}$Sb, we calculate the $(n,\gamma)$ rate and the MACS for $^{126}$Sb, the latter shown in Fig.~\ref{fig:MACSs}. 
This was done using TALYS~\cite{TALYS, TALYS2}. 
By using each NLD-GSF pair as input, we propagate both the statistical and systematic uncertainties of the NLD and the GSF by letting the resulting MACS inherit the $\chi^2$ score of the pair. 
From this, the uncertainty 
was found for each energy bin by graphically checking where the $\chi^2+1$ line would cross the parabola, similarly to what done before with the NLD and the GSF. 

The experimentally constrained MACS is compared to different libraries such as JINA REACLIB, TENDL-19, ENDF/B-VIII.0, and BRUSLIB, together with the span of all TALYS predictions available from each theoretical NLD and GSF model combination, including both microscopic and macroscopic models (light yellow band in Fig.~\ref{fig:MACSs}). 
We  see that the MACS of both TENDL-19 and JINA REACLIB are inside the $1\sigma$ confidence band and the same is true for the BRUSLIB library. 
All of those libraries are compatible with our estimated MACS.
However, the ENDF/B-VIII.0 library predicts a much higher rate (outside the experimental $2\sigma$ confidence), although it is still within the TALYS uncertainty band.
It is not clear why ENDF/B-VIII.0 predicts a much higher MACS than the others, but probably it is due to significant differences in the input NLD and GSF used for the evaluation of the MACS. 

The large variations in the NLD and GSF models are demonstrated in Figs.\ref{fig:nld} and \ref{fig:gsf}. 
The actual input models used in the libraries are not necessarily transparent, except for the BRUSLIB library which consequently uses the \texttt{ldmodel}~5 and \texttt{strength}~4~\cite{BRUSLIB}. Therefore, it is hard to explain why some of the library MACS are within the $1\sigma$ band of the present work and some are not.
To be able to conclude whether the $i$ process can explain  abundance observations, one needs to know the uncertainty in the $(n,\gamma)$ rates of the nuclei involved in the \textit{i} process.
Moreover, the abundance sensitivity to nuclear input is often evaluated by varying the $(n,\gamma)$ rates within some range. 
Unless known experimentally, the range might be determined from the variation in theoretical predictions using different NLD and GSF models,  (see, \textit{e.g.,} Refs.~\cite{Denissenkov2017, Cote2018, McKay}), or by vary the rates of a library (such as JINA REACLIB) by a fixed factor~(see \textit{e.g.} Ref.~\cite{Surman2014}).
Both these methods suffer from the problem that the models themselves usually do not provide parameter uncertainties. 
For a given rate, the uncertainty range might be too large, but also skewed, as the theoretical predictions are not necessarily normally distributed about the ``true'' value.
Therefore, it is of utmost importance to (\textit{i}) obtain as much experimental information as possible for nuclei relevant to the $i$ process, and (\textit{ii}) develop models of the NLD and GSF that are able to grasp the underlying physics, and at the same time provide reasonable estimates of existing experimental data.

\section{Summary and outlook}
\label{secSummary}
This work presents the measurement of the $^{124}$Sn($\alpha,p\gamma$)$^{127}$Sb reaction. 
We have experimentally extracted the NLD and GSF of $^{127}$Sb.
These quantities have allowed us to estimate the Maxwellian-averaged cross section for the $^{126}$Sb($n,\gamma$)$^{127}$Sb reaction, which is of interest for $i$ process network calculations. 
The resulting MACS is in agreement with the estimates from the JINA REACLIB, BRUSLIB and TENDL libraries.
In contrast, a significant discrepancy was found with the ENDF/B-VIII.0 library. 

It has been found that the GSF of $^{127}$Sb displays an upbend and a pygmy-like resonance at about $E_\gamma = 7$ MeV. 
By fitting models to the data, we have estimated that the strength in the PDR region corresponds to about 0.7-0.9\% of the Thomas-Reiche-Kuhn energy-weighted sum rule. 
A small peaklike structure was observed at about $E_\gamma = 4$ MeV, which is difficult to explain with theoretical models.
More precise measurements in this area, together with data of the GSF below 3 MeV and above the neutron-separation energy, would be desirable to better understand the behavior of these structures. 

It is  our hope that our data might inspire future developments of better theoretical  models for the GSF.
The impact of the data-constrained $(n,\gamma)^{127}$Sb MACS on final \textit{i} (and possibly \textit{r}) process abundances will be addressed  in a future work.

\section{Acknowledgments}
We would like to thank Pawel Sobas, Victor Modamio and Jon C. Wikne at the Oslo Cyclotron Laboratory for operating the cyclotron and providing excellent experimental conditions, and Fabio Zeiser for taking shifts during the experiment. 
A.C.L. gratefully acknowledges funding from the Research Council of Norway, Project No.~316116. 
The calculations were performed on resources provided by Sigma2, the National Infrastructure for High Performance Computing and Data Storage in Norway (using ``Saga,'' on Project No.~NN9464K). 
V.W.I., A. G., and S.S. gratefully acknowledge financial support from the Research Council of Norway, Project No.~263030. 
E.F.M. acknowledges support from the INTPART program from the Research Council of Norway, Project No.~310094. 
\bibliography{article127Sb}

\end{document}